\documentclass[preprint,12pt,numbers]{elsarticle}

\usepackage{amssymb}
\usepackage{algorithm}
\usepackage{algorithmic}
\usepackage{setspace}
\usepackage{threeparttable}
\usepackage{amsmath}
\usepackage{multirow}
\usepackage{subfigure}
\usepackage{bm}
\usepackage{color}
\usepackage{amssymb}
\usepackage{url}
\usepackage{adjustbox}
\usepackage{booktabs}
\usepackage{graphicx}
\usepackage{subfigure}

\newcommand{\eg}{\textit{e}.\textit{g}.}
\newcommand{\ie}{\textit{i}.\textit{e}.}
\newcommand{\et}{\textit{e}\textit{t} \textit{a}\textit{l}.}
\biboptions{sort&compress}

\journal{Journal of Pattern Recognition Templates}

\begin{document}
\doublespacing
\begin{frontmatter}



\title{Max360IQ: Blind Omnidirectional Image Quality Assessment with Multi-axis Attention}

% \title{Revisiting the Robustness of Spatio-temporal Modeling in Video Quality Assessment}


\author[a]{Jiebin Yan}
\author[a]{Ziwen Tan}
\author[a]{Yuming Fang*}
\author[a]{Jiale Rao}
\author[a]{Yifan Zuo}

\address{a School of Information Technology, Jiangxi University of Finance and Economics, Nanchang, China}
% \address{b SANY Heavy Industry Co., Ltd., Beijing, China}

% \cortext[1]{Corresponding author}

% \author[a]{Jiebin Yan*}
% \author[a]{Lei Wu}
% \author[a]{Wenhui Jiang}
% \author[b]{Chuanlin Liu}
% \author[c]{Fei Shen}

% \address{a School of Information Technology, Jiangxi University of Finance and Economics, Nanchang, China}
% \address{b Sichuan DOOV Intelligent Cloud Valley College.，Ltd., Yibin, China}
% \address{c SANY Heavy Industry Co., Ltd., Beijing, China}

% \cortext[1]{Corresponding author}

% \affiliation{organization={},%Department and Organization

\begin{abstract}
Omnidirectional image, also called 360-degree image, is able to capture the entire 360-degree scene, thereby providing more realistic immersive feelings for users than general 2D image and stereoscopic image. Meanwhile, this feature brings great challenges to measuring the perceptual quality of omnidirectional images, which is closely related to users' quality of experience, especially when the omnidirectional images suffer from non-uniform distortion. In this paper, we propose a novel and effective blind omnidirectional image quality assessment (BOIQA) model with multi-axis attention (Max360IQ), which can proficiently measure not only the quality of uniformly distorted omnidirectional images but also the quality of non-uniformly distorted omnidirectional images. Specifically, the proposed Max360IQ is mainly composed of a backbone with stacked multi-axis attention modules for capturing both global and local spatial interactions of extracted viewports, a multi-scale feature integration (MSFI) module to fuse multi-scale features and a quality regression module with deep semantic guidance for predicting the quality of omnidirectional images. Experimental results demonstrate that the proposed Max360IQ outperforms the state-of-the-art Assessor360 by 3.6\% in terms of SRCC on the JUFE database with non-uniform distortion, and gains improvement of 0.4\% and 0.8\% in terms of SRCC on the OIQA and CVIQ databases, respectively. The source code is available at \url{https://github.com/WenJuing/Max360IQ}.
\end{abstract}


  % \textcolor{red}{to deal with the uniform and non-uniform distortion simultaneously, we employ two sampling approaches to represent the OIs and design an optional component in the proposed model suited for different distortion situations.} 


\begin{highlights}
\item We propose a BOIQA model named Max360IQ based on stacked multi-axis attention modules, which is able to capture global degradation and local distortion, enabling us to address both non-uniform and uniform distortion.

\item We design a MSFI module to adaptively obtain the multi-scale quality-aware features and a Deep Semantic Guidance (DSG) module for better fusing the perceptual features of viewports.

\item The experiments show that the proposed Max360IQ obtains promising performance and outperforms other state-of-the-art methods. 
\end{highlights}

\begin{keyword}

% Moreover, we also study the influence of the number of input viewports on model performance and attain the balance of efficiency and effectiveness.


Omnidirectional images\sep perceptual quality assessment \sep multi-axis attention
\end{keyword}

\end{frontmatter}


\section{Introduction}
\label{sec:introduction}

Omnidirectional image (OI) is one of the important mediums of virtual reality and can provide an immersive experience for users by feat of Head Mounted Displays (HMDs). Ensuring good quality of experience (QoE) for users is one of the basic requirements for application systems. However, during the procedure of acquisition, transmission, processing, storage, and \emph{etc}, OIs may suffer from distortion and thus their quality would be far from being satisfactory, which significantly degrades the QoE of users~\cite{wang2017begin}. Thus, accurately estimating the quality of OIs is of great importance for system optimization, algorithm optimization~\cite{fang2021superpixel,zhang2023data,zhou2023surroundnet}, and helping suppliers automatically select high-quality OIs from massive visual data, and testing the performance of omnidirectional cameras and equipment. Generally, omnidirectional image quality assessment (OIQA) models can be divided into three categories, including Full-Reference OIQA (FR-OIQA), Reduced-Reference OIQA (RR-OIQA), and No-reference/Blind OIQA (BOIQA). The former two types of model need to use full and partial reference information when being deployed, respectively; the last type of models can evaluate image quality without access to reference information, and thus are more practical than the former two types of model.

\begin{figure}
    \centering
    \subfigure{
        \includegraphics[width=0.64\textwidth]{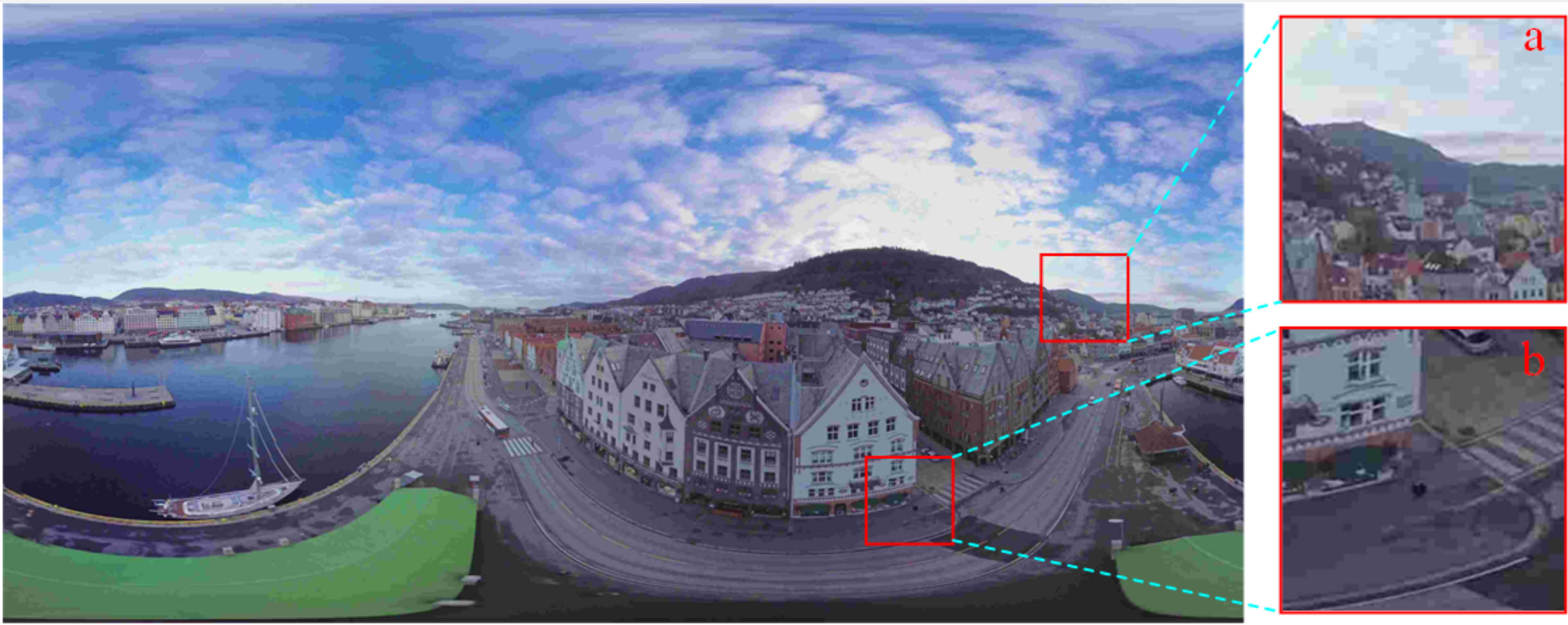}
    }
    \subfigure{
        \includegraphics[width=0.64\textwidth]{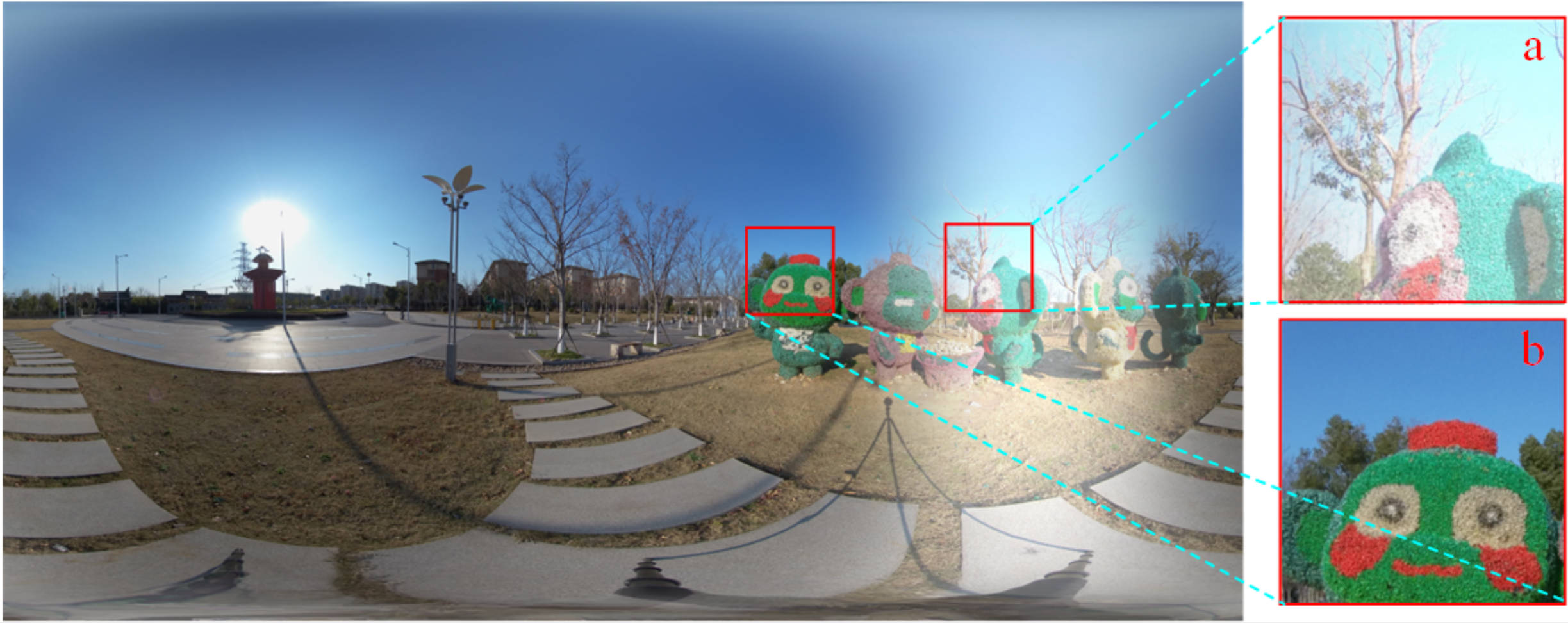}
    }
    \caption{The visual examples of uniformly distorted OI and non-uniformly distorted OI. The top row shows an OI suffered from uniform distortion from the CVIQ database~\cite{sun2018large}, while the bottom row shows an OI with non-uniform distortion from the JUFE database~\cite{fang2022perceptual}.}
    \label{fig:dis_example}
\end{figure}

In previous years, some patch-based BOIQA methods~\cite{kim2019deep,lim2018VR} have been proposed, which divide an OI in Equirectangular Projection (ERP) format into equally sized patches and use a convolutional neural network (CNN) to extract quality features of those patches, and then regress them to a quality score. However, these approaches overlook the deformation at the poles of OIs caused by projection stretching, resulting in inconsistent representations of patches and discrepancies with users' visual perception. In comparison, the viewport-based methods~\cite{sun2019mc360iqa,zhou2021omnidirectional,fang2022perceptual,jabbari2023st360iq,wu2024assessor360} are able to solve the above problems, since the \textit{viewport} can provide a more reasonable representation of human perception. Although these viewport-based methods have shown promising results on the uniformly distorted OI quality databases such as OIQA~\cite{duan2018perceptual} and CVIQ~\cite{sun2018large}, their performance on the non-uniformly distorted OIs is sub-optimal. The visual examples of uniformly distorted OIs and non-uniformly distorted OIs are shown in Fig.~\ref{fig:dis_example}.

In this paper, we design a unified OIQA model for effectively capturing both non-uniform and uniform distortion. Specifically, we first extract the suitable viewport sequences for each OI. Furthermore, we utilize stacked multi-axis attention modules~\cite{tu2022maxvit} to capture both global and local spatial interactions. Inspired by the effectiveness of multi-scale features in OIQA models~\cite{fu2022adaptive,zhang2022no}, a multi-scale feature integration (MSFI) module  is employed to adaptively incorporate multi-scale features of viewports. The Gate Recurrent Units (GRUs) are adopted to special for processing non-uniform distortion. Finally, we design a deep semantic guided quality regression module for mapping the extracted features to a quality score.

In summary, our contributions are three folds:

\begin{itemize}
\item We propose a BOIQA model named Max360IQ based on stacked multi-axis attention modules, which is able to capture global degradation and local distortion, enabling us to address both non-uniform and uniform distortion.
\item We design a MSFI module to adaptively obtain the multi-scale quality-aware features and a Deep Semantic Guidance (DSG) module for better fusing the perceptual features of viewports.
\item The experiments show that the proposed Max360IQ obtains promising performance and outperforms other state-of-the-art methods.
\end{itemize}

% ********************************************************

\section{Related work}
\label{sec:rw}

In this section, we first briefly introduce blind image quality assessment (BIQA) models for general 2D images. Then, we describe OIQA models in detail.

\subsection{BIQA Models}

Most of the early BIQA models (also called traditional models) follow the same paradigm, \ie, extract hand-crafted features based on prior knowledge and predict perceptual quality using a shallow machine learning algorithm (\eg, support vector regression)~\cite{fang2017no} or a distance metric~\cite{mittal2012making}. This is the mainstream framework before the deep learning era. Limited by the representation ability of hand-crafted features and the learning ability of the shallow machine learning algorithms, the performance of traditional models has been largely surpassed by deep learning-based models.  

To the best of our knowledge, Kang~\et~\cite{kang2014convolutional} presented a pioneering study on IQA using a CNN, and it can be trained end-to-end. After that, many CNN-based BIQA models have been successively proposed ~\cite{kim2017deep,zhang2018blind,gu2019blind,su2020blindly}, where most of them are adapted from the CNNs for image classification~\cite{kim2017deep,zhang2018blind,gu2019blind,su2020blindly}. In addition, some CNN-based BIQA models target solving the data-scarcity problem, since CNN is data-hungry while the amount of public image quality data is very small. For example, this two studies~\cite {liu2017rankiqa, ma2017dipiq} introduce pair-wise ranking learning to solve the data problem, where the discriminable image pairs with preference labels can be generated automatically without any constraint. Some works focus on improving the model's generalization ability by integrating advanced computer vision techniques, such as meta-learning~\cite {zhu2020metaiqa}, active learning~\cite{wang2021active}, and \emph{etc}. Besides, Su~\et~\cite{su2023distortion} proposed an alternative solution to improve model generalizability, where it learns an image distortion manifold to capture common degradation patterns, and the quality prediction of an image can be obtained by projecting the image to the distortion manifold.

\subsection{OIQA Models}

Due to the substantial difference between the data format of OIs and general 2D images, the 2D-IQA models can not be directly used to estimate the quality of OIs. Some researchers made attempts to adapt the 2D-IQA models to solve the OIQA issue. Yu~\et~\cite{yu2015framework} proposed S-PSNR, which projects an OI to the ERP format and calculates the PSNR value between the reference OI and distorted OI. Sun~\et~\cite{sun2017weighted} proposed WS-PSNR, which assigns different weights to the pixels at different positions. Zakharchenko~\et~\cite{zakharchenko2016quality} proposed CPP-PSNR, which maps the OIs to the Craster Parabolic Projection (CPP) format and then calculates the PSNR. Besides, some researchers tried to adapt SSIM~\cite{wang2004image} to design OIQA models, such as S-SSIM~\cite{chen2018spherical} and WS-SSIM~\cite{zhou2018weighted}.

For the BOIQA models, they can be divided into three types, including ERP-based, other projection format-based, and viewport-based. The main idea of the first type of BOIQA model is to directly extract features from the ERP image. Yang~\et~\cite{yang2021spatial} proposed a spatial attention-based model named SAP-net, which fuses the error map generated by the difference between the impaired image and the enhanced image into the backbone to 
explicitly emphasize the objective degradation. Sendjasni~\et~\cite{sendjasni2023attention} proposed an attention-aware patch-based OIQA model that considers the exploration behavior and latitude-based selection in the sampling process. Kim~\et~\cite{kim2017deep} proposed an OIQA method based on adversarial neural networks which divides the ERP image into uniform and non-overlapping patches, and uses the feature extracted from these patches to predict quality score. The second type is mainly to overcome the obvious stretching and deformation of ERP images at the poles, and some researchers tried to project OIs to other formats. Jiang~\et~\cite{jiang2021cubemap} proposed to extract features from the six projected images in the CMP format. Sun~\et~\cite{sun2019mc360iqa} proposed an OIQA model using a multi-channel CNN named MC360IQA to simulate the real viewing process of the subjects. This model projects each OI into six viewports and uses a multi-channel CNN to extract features, then the extracted features are fused and used to predict the quality score. 
Xu~\et~\cite{xu2020blind} proposed a novel viewport-oriented graph neural network (GNN) named VGCN, which considers local viewport quality and global image quality simultaneously. Yang~\et~\cite{yang2022tvformer} proposed a novel Transformer-based model named TVFormer, which generates the viewport sequence by a trajectory-aware module and predicts the perceptual quality of OIs in a manner of video quality assessment (VQA). Fang~\et~\cite{fang2022perceptual} proposed a BOIQA model that incorporates the subject viewing conditions to maintain consistency with human viewing behavior. Wu~\et~\cite{wu2024assessor360} proposed a multi-sequence network called Assessor360 which generates multiple pseudo viewport sequences as the inputs.

\begin{figure*}[]
\centering
\includegraphics[width=1\linewidth]{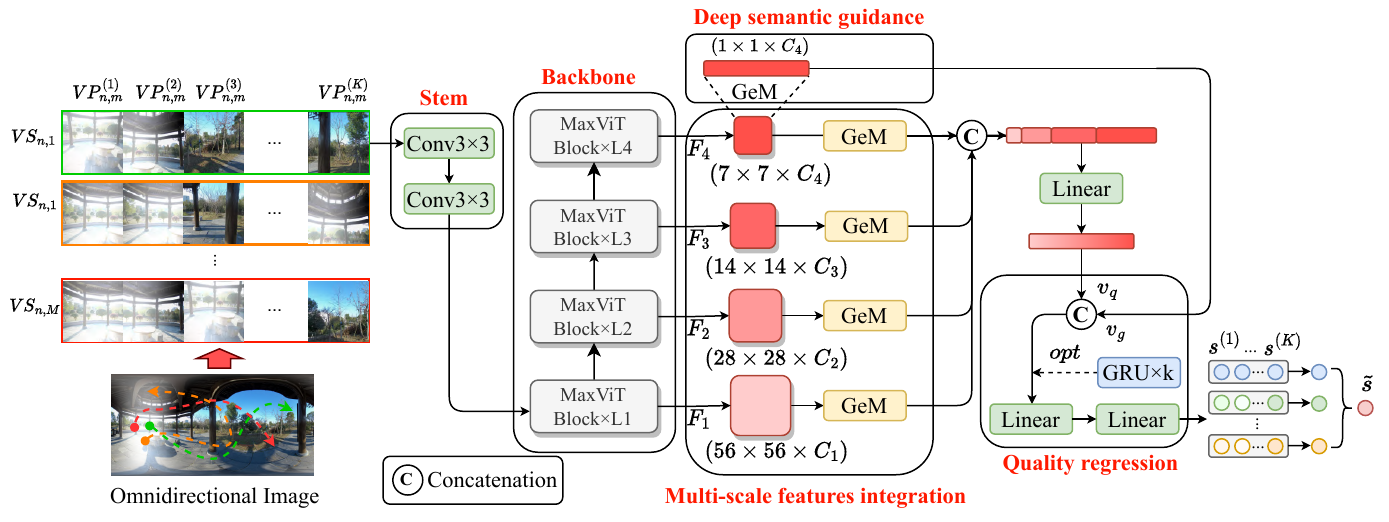}
\caption{The architecture of our proposed Max360IQ. It mainly consists of three parts: a backbone, a multi-scale feature integration (MSFI) module, and a quality regression (QR) module. Note that the GRUs component in Max360IQ is optional for optimal performance in different scenarios, \ie, non-uniformly and uniformly distorted OIs.}
\label{fig:max360iq}
\end{figure*} 

% ********************************************************

\section{Proposed Method}
\label{sec:pm}

This section provides a detailed introduction to the proposed Max360IQ, whose framework is shown in Fig.~\ref{fig:max360iq}. We first extract the viewports of each OI, then employ a backbone with stacked multi-axis attention modules to extract local and global interactions for each viewport. The MSFI module is used to adaptively fuse features from different scales and the optional GRUs are used to process the non-uniform distortion. Finally, the fused multi-scale features and deep semantic features are used as the input of the quality prediction module.

\subsection{Viewport Extraction}

\begin{figure}[htbp]
    \centering

    \subfigure[Good-5s]{
        \includegraphics[width=0.22\textwidth]{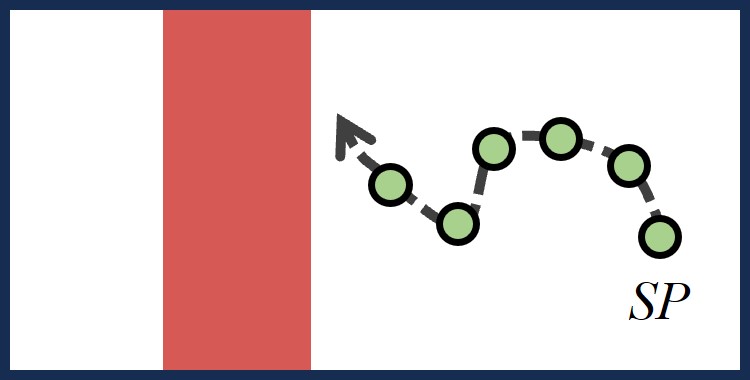}
    }
    \subfigure[Bad-5s]{
        \includegraphics[width=0.22\textwidth]{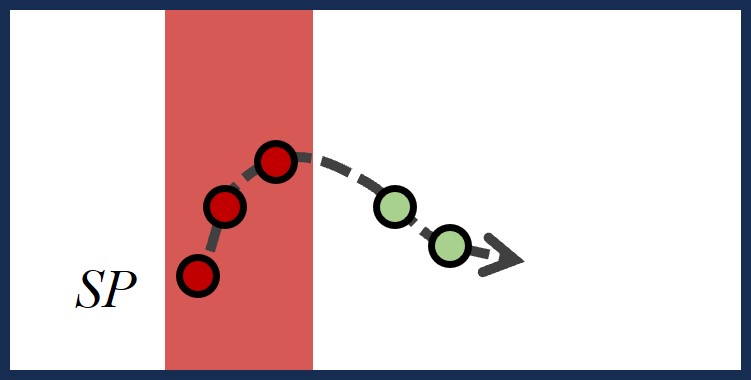}
    }
    \subfigure[Good-15s]{
        \includegraphics[width=0.22\textwidth]{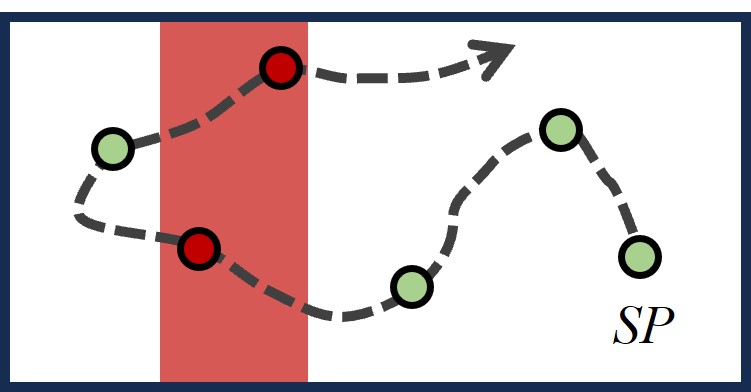}
    }
    \subfigure[Bad-15s]{
        \includegraphics[width=0.22\textwidth]{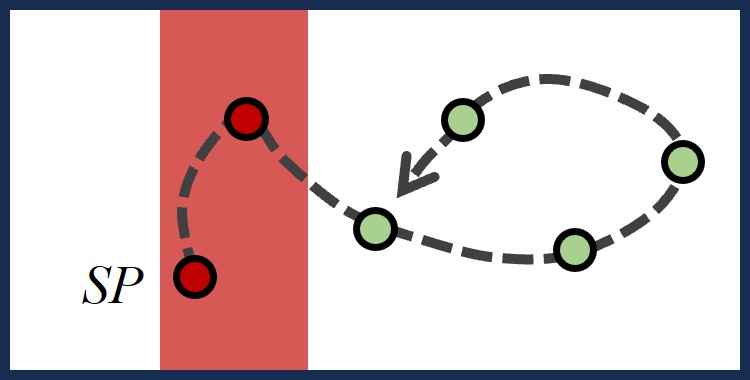}
    }
    \caption{The illustration of the viewport extraction of four conditions. Note that the $SP$ denotes the starting point of scanpath and the region marked in red denotes low quality.}
    \label{fig:four_condition}
\end{figure}

For the non-uniformly distorted OIs, since the viewing conditions, \ie, starting point and exploration time, are of great importance in perceiving their quality~\cite{fang2022perceptual}, we extract the viewport sequences according to the viewing conditions. Specifically, we first consider the starting point which is relative to the initial impression in the viewing process, the good starting point and the bad starting point are considered. Note that the good starting point means that the region where each user starts viewing is with high quality, and the bad starting point denotes that the region is with worse quality when each user starts viewing. Besides, we also consider the exploration times, which are set to 5s and 15s, the longer time means the possible wider viewing range. Thus, there are four viewing conditions: Good-5s, Bad-5s, Good-15s, Bad-15s. For each viewing condition, we uniformly sample $K$ (the default value is set to 7) viewports according to the default 300 corresponding coordinates of scanpath offered by the JUFE database~\cite{fang2022perceptual}, where each viewport sequence corresponds to a subjective quality score, the visual examples are shown in Fig.~\ref{fig:four_condition}. For testing the performance of the proposed Max360IQ on predicting the quality of uniformly distorted OIs, \ie, the OIQA~\cite{duan2018perceptual} and CVIQ~\cite{sun2018large} databases, we simply extract several viewports equidistantly along the equator of each ERP image to consist of viewport sequences. The influence of the number of extracted viewports will be discussed in Section~\ref{sec:exper}. 

In summary, given a database with $N$ OIs denoted by $\{OI_n\}_{n=1}^N$, the training data consists of $D=\{VS_{n, m}\}_{m=1}^M$, where $M$ represents the number of viewport sequences of $n$-th OI. $VS_{n,m}=\{VP_{n,m}^{(k)}\}_{k=1}^K$ represents a viewport sequence, where $VP_{n,m}^{(k)}$ denotes the $k$-th viewport in the $m$-th viewport sequence of the $n$-th OI. Before inputting the viewports in the backbone, each viewport sequence will be pre-processed by the Stem module, which is used to adjust the viewports' shape, reduce complexity, and extract initial features.

% User behavior significantly impacts the perceived quality of OIs~\cite{fang2022perceptual}. To align with this observation, we extract the actual viewport sequences by considering the starting point, viewing time, and scanpath for each OIs. Besides, building upon the findings of Assessor360, excessive viewports can introduce redundant information that disrupts network training. To balance the effectiveness and efficiency, for the 15-second sequences, we uniformly sample wholly 300 key points from each scanpath and reduce them to 7 key points. For the 5-second sequences, we perform downsampling by selecting only the first 100 key points. Subsequently, viewport sequences can be generated by extracting the viewports from the corresponding coordinates in chronological order. As a result, there are four conditions in viewport sequences: the good starting point for 5/15 seconds (g5/g15), and the bad starting point for 5/15 seconds (b5/b15). Each condition's viewport sequence corresponds to a specific Mean Opinion Score (MOS) which can be obtained in the JUFE database. 

\subsection{Backbone}
In our implementation, we adopt a sequence of MaxViT blocks~\cite{tu2022maxvit} as the backbone, which consists of four stages and each stage contains several identical MaxViT blocks. These blocks are responsible for extracting quality features of each viewport considering both local details and global semantics with linear complexity, making the feature extraction process efficient. The size of features from the last layer of each stage is halved by a max pooling layer, and the dimensions of these features in the respective stages are $\{C_1, C_2, C_3, C_4\}$. For the $k$-th viewport, the above procedure can be formulated as: 
\begin{equation}
{F_4}=Blocks(Stem(VP;\theta_s);\theta_b),
\end{equation}
where the $\theta_s$ and $\theta_b$ denote the parameters for Stem and MaxViT blocks, respectively. $F_4$ denotes the feature map from the last layer of the backbone. Here, $m$, $n$, and $k$ are omitted for simplicity.

\subsection{Multi-scale Features Integration}
Considering the multi-scale characteristic of the human visual system (HVS)~\cite{su2020blindly,wu2020end,yan2020blind}, we design a MSFI module to extract multi-scale quality-aware features of each OI. Besides, we note that the multi-scale feature integration operation has been widely used in OIQA models in a fixed manner~\cite{fu2022adaptive,zhang2022no}, however, the extraction and integration of multi-scale features could be a dynamic process due to the complexity of subjective perception. Thus, we employ four Generalized-Mean (GeM) pooling layers~\cite{radenovic2018fine} to adaptively adjust the features in each scale as well as reduce the computational load. Specifically, we denote the output feature maps of the four stages as $\{F_1, F_2, F_3, F_4\}$, and four GeM pooling layers are employed to further process these feature maps.

The GeM pooling operation to the $i$-th $F$ can be formulated as:
\begin{equation}
f_i=GeM_i(F_i),i=1, \cdots,4,
\end{equation}
where $F_i$ denotes the feature map from the last layer of the $i$-th stage, and the GeM pooling is defined as:
\begin{equation}
f_i = \left(\frac{1}{\|F_i\|}\sum_{x_i \in F_i}x_i^{\rho}\right)^{\frac{1}{\rho}},
\end{equation}
where $x_i$ denotes the pixel value on the feature map $F_i$. Note that the pooling factor $\rho$ is learnable. The GeM pooling approaches to max pooling when $\rho$ close to 
infinity, and approaches to average pooling when $\rho$ close to 1. The outputs of the GeM pooling operations are concatenated and then fused by a fully-connected layer, resulting in a multi-scale quality-aware feature vector $v_q$. The operation is as follows:
\begin{equation}
v_q = FC\left(f_1\cup f_2\cup f_3\cup f_4\right),
\end{equation}
where $FC$ denotes the fully connection (FC) layer; $\cup$ denotes the concatenation operation.
\subsection{Quality Regression}
\emph{1) Deep Semantic Guidance}: Apart from multi-scale features, we also use deep semantic features to assist quality prediction tasks. We firstly use a GeM pooling to pre-process $F_4$, and then concatenate it with the fused multi-scale features $v_q$. The procedure can be formulated as follows:
\begin{equation}
v_g = GeM(F_4),
\end{equation}
\begin{equation}
\tilde{v} = v_q \cup v_g.
\end{equation}

\emph{2) Tackling Recency Effect}: To simulate the recency effect~\cite{hands2001recency}, \textit{i.e.}, recent content has a greater impact on overall impression, we employ several GRUs to capture temporal dependency. Subsequently, the predicted quality score  $s_{n,m}^{(k)}$ of $VP_{n,m}^{(k)}$ is obtained through regression using two fully-connected layers. The final quality score $\tilde{s}$ for the input OI is derived by two average operations. The regression procedure can be formulated as follows:
\begin{equation}
[s_{n,m}^{(1)}, \cdots, s_{n,m}^{(K)}] = FC_2(FC_1(GRUs([\tilde{v}_{n,m}^{(1)}, \cdots, \tilde{v}_{n,m}^{(K)}]))).
\end{equation}
Subsequently, the quality of $m$-th viewport sequence of $n$-th OI can be obtained by:
\begin{equation}
\label{eq:s1}
\tilde{s}_{n,m} = \frac{1}{K}\sum_{k=1}^K s_{n,m}^{(k)},
\end{equation}
where $K$ denotes the number of viewports in one viewport sequence. The final score of the $n$-th OI can be obtained by:
\begin{equation}
\label{eq:s2}
\tilde{s}_{n} = \frac{1}{M}\sum_{m=1}^M \tilde{s}_{n,m},
\end{equation}
where $M$ denotes the number of viewport sequences of the OI.

\emph{3) Loss function}: In this study, we adopt the \textit{Norm-in-Norm}~\cite{li2020norm} loss function to optimize the proposed model, since it enables faster convergence and allows more robust parameters compared to the mean absolute error (MAE) and mean square error (MSE) loss. Given the predicted quality scores $\bm{\tilde s}=(\tilde s_1,\cdots,\tilde s_N)$ and the subjective quality scores $\bm{\hat s}=(\hat s_1,\cdots,\hat s_N)$, the definition of loss is as follows:

\begin{equation}
L(\tilde s_n,\hat s_n)=\frac{1}{\epsilon}\sum_{n=1}^N{\|\tilde q_n-\hat q_n\|}^p,
\end{equation}
where $\epsilon$ is a normalization factor; $p$ is a hyper-parameter; $\tilde s_n$ and $\hat s_n$ denote the predicted quality score and the subjective score of the $n$-th OI in the training set; $\tilde q_n$ and $\hat q_n$ denote the normalization quality score, respectively, and they are defined as follows:

\begin{equation}
\tilde q_n=\frac{\tilde s_n-\tilde \mu}{\tilde \sigma},n=1,\cdots,N,
\end{equation}
\begin{equation}
\hat q_n=\frac{\hat s_n-\hat \mu}{\hat \sigma},n=1,\cdots,N,
\end{equation}
where $\tilde \mu$ and $\hat \mu$ are the mean of the predicted quality score and the subjective score, respectively; $\tilde \sigma$ and $\hat \sigma$ are the norm of the predicted quality score and the subjective score, respectively. Their definitions are as follows:
\begin{equation}
\tilde \mu=\frac{1}{N}\sum_{n=1}^N \tilde s_n,\tilde \sigma=\left(\sum_{n=1}^N \|\tilde s_n-\tilde \mu\|^q\right)^{\frac{1}{q}},
\end{equation}
\begin{equation}
\hat \mu=\frac{1}{N}\sum_{n=1}^N \hat s_n,\hat \sigma=\left(\sum_{n=1}^N \|\hat s_n-\hat \mu\|^q\right)^{\frac{1}{q}},
\end{equation}
where $q\geq 1$ is a hyper-parameter.

For ease of understanding, Tab.~\ref{tab:abbr} provides the abbreviations used in the paper along with their corresponding meanings.

\begin{table}[!htbp]
    \centering
    \caption{The summary of abbreviations involved in this paper.}
    \label{tab:abbr}
    \begin{adjustbox}{max width=1\textwidth}
    \begin{tabular}{c|c|c|c}
        \toprule
        Abbreviations & Description of the abbreviations & Abbreviations & Description of the abbreviations \\
        \midrule
        OI & Omnidirectional image  & SRCC& Spearman’s rank order correlation coefficient\\
        BIQA&Blind image quality 
        assessment&GeM&Generalized-mean\\
        VQA&Video quality assessment&$N$ &The number of omnidirectional images \\
        OIQA & Omnidirectional image quality assessment& $M$ & The number of viewport sequences\\
        FR-OIQA & Full-reference OIQA& $K$ & The number of viewports\\
        RR-OIQA&Reduced-reference OIQA & $C$& The dimensions of features\\
        BOIQA&Blind OIQA & $F$& Feature map\\

        ERP&Equirectangular projection & VP& Viewport\\
        HVS&Human visual system& VS& Viewport sequence\\
        HMDs & Head mounted displays & $\mu$&Mean\\
        QoE & Quality of experience& $\sigma$& Norm\\

        CNN&Convolutional neural network & $\tilde s$&The final predicted quality score \\
        GNN&Graph neural network&$s$& The predicted quality score\\
        
        GRUs&Gate recurrent units & $\hat s$&The subjective quality score  \\
        DSG&Deep semantic guidance&$q$&Mormalization quality score\\
        MSFI & Multi-scale feature integration & $v_q$&Multi-scale quality-aware feature vector\\

        PLCC&Pearson’s linear correlation coefficient&$\theta$&Parameter\\
        \bottomrule
    \end{tabular}
    \end{adjustbox}
\end{table}
%-------------------------------------------------------------------------
\section{Experimental Results}
In this section, we introduce the details of experimental implementation and provide insightful analyses.

\subsection{Test Database}
To comprehensively evaluate the performance of Max360IQ, we compare it with other state-of-the-art methods on the JUFE~\cite{fang2022perceptual}, OIQA~\cite{duan2018perceptual} and CVIQ~\cite{sun2018large} databases in our experiments. The details of the three databases are as follows.

\emph{1) JUFE}: As the first large database for studying the non-uniform distortion of OIs, JUFE includes 258 original OIs and 1,032 non-uniformly distorted OIs, where all images have the 8K resolution. The distorted images are generated by adding synthesized distortions, including Gaussian blur, Gaussian noise, brightness discontinuity, and stitching, and each type of distortion is with three levels and randomly applied to one of the six lenses in the OIs. This database also records eye movement data.

\emph{2) OIQA}: It consists of 16 original OIs and 320 uniformly distorted OIs, each original OI enjoys a wonderful color and composition with high resolution (from 11K to 13K). The distorted images are degraded by four types of distortions, including Gaussian blur, Gaussian white noise, JPEG compression and JPEG2000 compression, and each type of distortion is with five levels.

\emph{3) CVIQ}: To study the impact of compression on the perceptive quality of OIs, CVIQ contains 16 original OIs and 528 uniformly distorted OIs, where all images have the 4K resolution. There are totally three types of compression distortions, including JPEG, H.264/AVC, and H.265/HEVC, and each type of distortion is with 11 levels.

\subsection{Experimental Methodology}
\emph{1) Training Details}: We split the each database into 80\% for training and 20\% for testing, respectively. Our model is trained on a computer with an NVIDIA GeForce GTX 3090 Ti GPU. We utilize the Adam optimizer~\cite{kingma2014adam} to optimize the proposed Max360IQ with an initial learning rate of $1\times 10^{-4}$ and weight decay of $5\times 10^{-4}$. During training, we set the batch size to 16, the training epochs to 10 for the JUFE database and 150 for the OIQA and CVIQ databases. Besides, the dropout rate is set to 0.1 for the JUFE database and 0.5 for the OIQA and CVIQ databases. In the testing phase, each test OI's final quality score is calculated by averaging the predictions of all corresponding viewports as described in \eqref{eq:s1} and \eqref{eq:s2}.

\emph{2) Evaluation Metrics}:
We employ Pearson's linear correlation coefficient (PLCC), Spearman's rank order correlation coefficient (SRCC) and root mean square error (RMSE) to test the compared methods. 

PLCC is used to measure the accuracy of the model's predictions, which is defined as:
\begin{equation}
PLCC=\frac{\sum_{n=1}^N(\hat s_n-\mu_{\hat s})(\tilde s_n-\mu_{\tilde s})}{\sqrt{\sum_{n=1}^N(\hat s_n-\mu_{\hat s})^2}\sqrt{\sum_{n=1}^N(\tilde s_n-\mu_{\tilde s})^2}},
\end{equation}
where $N$ denotes the number of testing OIs; $\hat s_n$ and $\tilde s_n$ denote the subjective quality score and the mapped predicted quality score of the $n$-th OI; $\mu_{\hat s}$ and $\mu_{\tilde s}$ denote the mean of the $\hat s$ and $\tilde s$, respectively. 

SRCC assesses the monotonicity consistency between the model's predicted values and the ground truth, which is defined as:
\begin{equation}
SRCC=1-\frac{6\sum_{n=1}^Nd_n^2}{N(N^2-1)},
\end{equation}
where $d_n$ denotes the sorting difference between subjective scores and predicted quality scores, it is defined as:
\begin{equation}
d_n=X_n-Y_n,
\end{equation}
where $X_n$ and $Y_n$ denote the sorting number of subjective score and predicted score in the database, respectively.

RMSE measures the error in the model's predictions, which is defined as:
\begin{equation}
RMSE=\sqrt{\frac{1}{N}\sum_{n=1}^N(\hat s_n-\tilde s_n)^2}.
\end{equation}
Note that higher PLCC and SRCC, and lower RMSE represent better results. As suggested in~\cite{video2000final}, we apply a five-parameter logistic function to obtain the mapped predicted scores before calculating these metrics, which is defined as:
\begin{equation}
f(x)=\theta_1(\frac{1}{2}-\frac{1}{1+e^{\theta_2(x-\theta_3)+\theta_4x+\theta_5}}),
\end{equation}
where $x$ denotes the predicted quality score; $\theta_1,\theta_2,\cdots,\theta_5$ denotes the fitting parameters.

\emph{3) Compared Methods}:
We compare the proposed Max360IQ with nine models, including four FR-OIQA models: S-PSNR~\cite{yu2015framework}, WS-PSNR~\cite{sun2017weighted}, CPP-PSNR~\cite{zakharchenko2016quality} and WS-SSIM~\cite{zhou2018weighted}, and five BOIQA methods: MC360IQA~\cite{sun2019mc360iqa}, VGCN~\cite{xu2020blind}, Fang22~\cite{fang2022perceptual}, ST360IQ~\cite{jabbari2023st360iq} and Assessor360~\cite{wu2024assessor360}. To ensure a fair comparison on the JUFE database, MC360IQA, ST360IQ, VGCN and Assessor360 are retrained using this database, and their results on the OIQA and CVIQ databases are directly taken from their original papers. Note that the default input of ST360IQ is a single viewport, thus it is trained on the data under each condition separately.

\subsection{Experimental Results}
\label{sec:exper}

The experimental results are shown in Tab.~\ref{tab:results1}. First, we can clearly observe that these FR-OIQA methods perform worse than other methods, and all of these FR-OIQA methods show poor performance on the JUFE database, which indicates that quality modeling of non-uniformly distorted OIs is much hard than that of uniformly distorted OIs. By contrast, these deep learning-based BOIQA models can automatically simulate the viewing behavior by end-to-end learning and therefore can obtain better performance than these FR-OIQA methods. Besides, we can find that ST360IQ shows poor performance on the JUFE dataset, despite its success on the OIQA and CVIQ datasets. The reason is that the model is trained using a single viewport as input, making it unable to capture non-uniform distortion. In contrast, the Max360IQ shows great performance improvement against the state-of-the-art methods on the JUFE database. Specifically, Max360IQ achieves superior performance over Assessor360 by 3.5\% in terms of PLCC and 3.6\% in terms of SRCC on the JUFE dataset. Moreover, the proposed Max360IQ also presents outstanding performance on the uniformly distorted OIs, \textit{i.e.}, OIQA and CVIQ databases, which has advantages on the OIQA database according to the improvement of 0.4\% in terms of SRCC, and the performance on the CVIQ database increased by 0.8\% in terms of PLCC and SRCC compared to the Assessor360. Note that Assessor360 outperforms our model in terms of PLCC on the OIQA database, whose possible reason is that its entropy-based sampling approach can obtain diverse viewport sequences, leading to exceptional performance in certain scenarios.

\begin{table*}[!htbp]
    \centering
    \caption{Performance comparison of state-of-the-art FR-OIQA and BOIQA methods on the JUFE, OIQA and CVIQ databases. The best results are marked in \textbf{bold}.}
    \label{tab:results1}
\begin{adjustbox}{max width=1\textwidth}
\begin{tabular}{c|c|c|c|c|c|c|c|c|c|c|c|c|c}
\toprule
Type & Metrics & \multicolumn{3}{c|}{JUFE} & \multicolumn{3}{c|}{OIQA} & \multicolumn{3}{c|}{CVIQ} & \multicolumn{3}{c}{Average}  \\
\midrule
  &  & PLCC & SRCC & RMSE & PLCC & SRCC & RMSE & PLCC & SRCC & RMSE & PLCC & SRCC & RMSE \\
\midrule
\multirow{4}{*}{FR-OIQA} & S-PSNR~\cite{yu2015framework} & 0.0869 & 0.0771 & 0.4075 & 0.5997 & 0.5399 & 1.6721 & 0.7083 & 0.6449 & 9.8564 & 0.4649 & 0.4199 & 3.9787\\
  & WS-PSNR~\cite{sun2017weighted} & 0.0895 & 0.0873 & 0.4074 & 0.5819 & 0.5263 & 1.6994 & 0.6729 & 0.6107 & 10.3283
 & 0.4445 & 0.4052 & 4.1650\\
  & CPP-PSNR~\cite{zakharchenko2016quality}  & 0.0885 & 0.0780 & 0.4075 & 0.5683 & 0.5149 & 1.7193
 & 0.6871 & 0.6265 & 10.1448 & 0.4478 & 0.4068 & 4.0905\\
  & WS-SSIM~\cite{zhou2018weighted} & 0.0757 & 0.0656 & 0.4082 & 0.5044 & 0.5032 & 1.8256 & 0.9293 & 0.9116 & 5.2626 & 0.4994 & 0.4909 & 2.4989\\
\midrule
\multirow{6}{*}{BOIQA} & MC360IQA~\cite{sun2019mc360iqa}  & 0.3901 & 0.3661 & 0.3528 & 0.9267 & 0.9139 & 0.7854 & 0.9429 & 0.9428 & 4.6506 & 0.5432 & 0.7409 & 1.9296 \\
  & VGCN~\cite{xu2020blind} & 0.4502 & 0.4264 & 0.3298 & 0.9584 & 0.9515 & 0.5967 & 0.9651 & 0.9639 & 3.6573 & 0.7912 & 0.7806 & 1.5279 \\
  & Fang22~\cite{fang2022perceptual} & 0.3158 & 0.2994 & 0.3665 & - & - & - & - & - & - & - & - & - \\
  & ST360IQ~\cite{jabbari2023st360iq} & 0.2176 & 0.2183 & 0.3359 & 0.9600 & 0.9700 & 0.5700 & 0.9800 & 0.9800 & 2.9800 & 0.7192 & 0.7228 & 1.2953 \\
  & Assessor360~\cite{wu2024assessor360} & 0.4619 & 0.4335 & 0.3323 & \textbf{0.9757} & 0.9661 & - & 0.9836 & 0.9801 & - & 0.8080 & 0.7990 & - \\
  & Max360IQ  & \textbf{0.4967} & \textbf{0.4696} & \textbf{0.3253} & 0.9699 & \textbf{0.9704} & \textbf{0.4959} & \textbf{0.9844} & \textbf{0.9809} & \textbf{2.3485} & \textbf{0.8164} & \textbf{0.8070} & \textbf{1.0566} \\
\bottomrule
\end{tabular}
\end{adjustbox}
\end{table*}

\begin{figure}[htbp]
    \centering
    
    \subfigure[S-PSNR]{
        \includegraphics[width=0.3\textwidth]{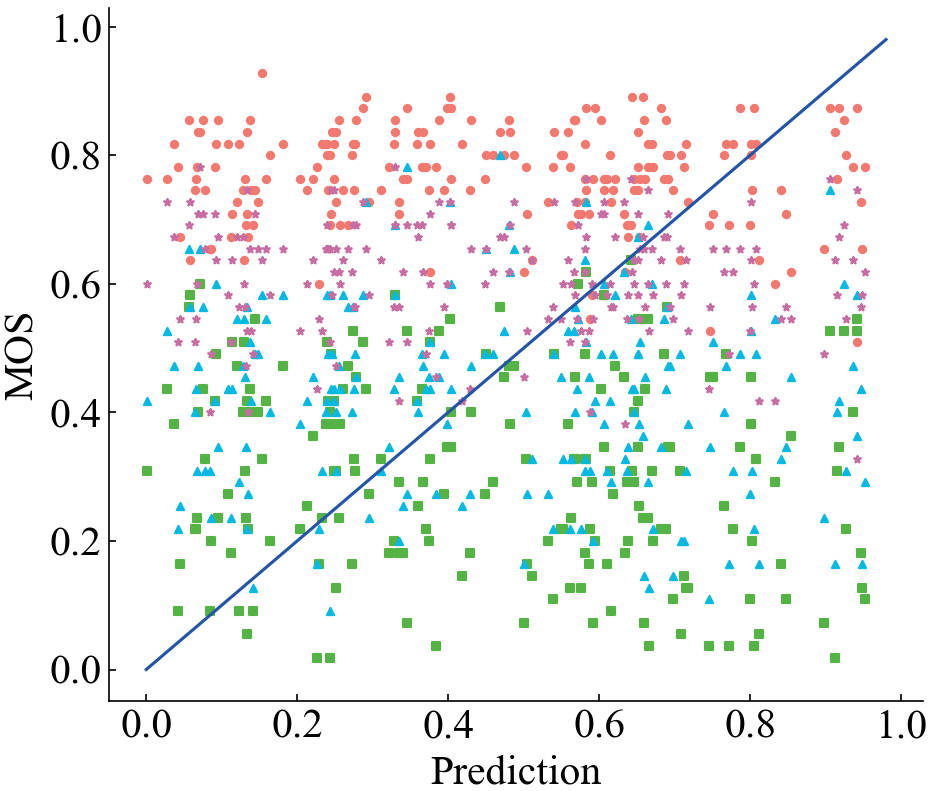}
    }
    \subfigure[WS-PSNR]{
        \includegraphics[width=0.3\textwidth]{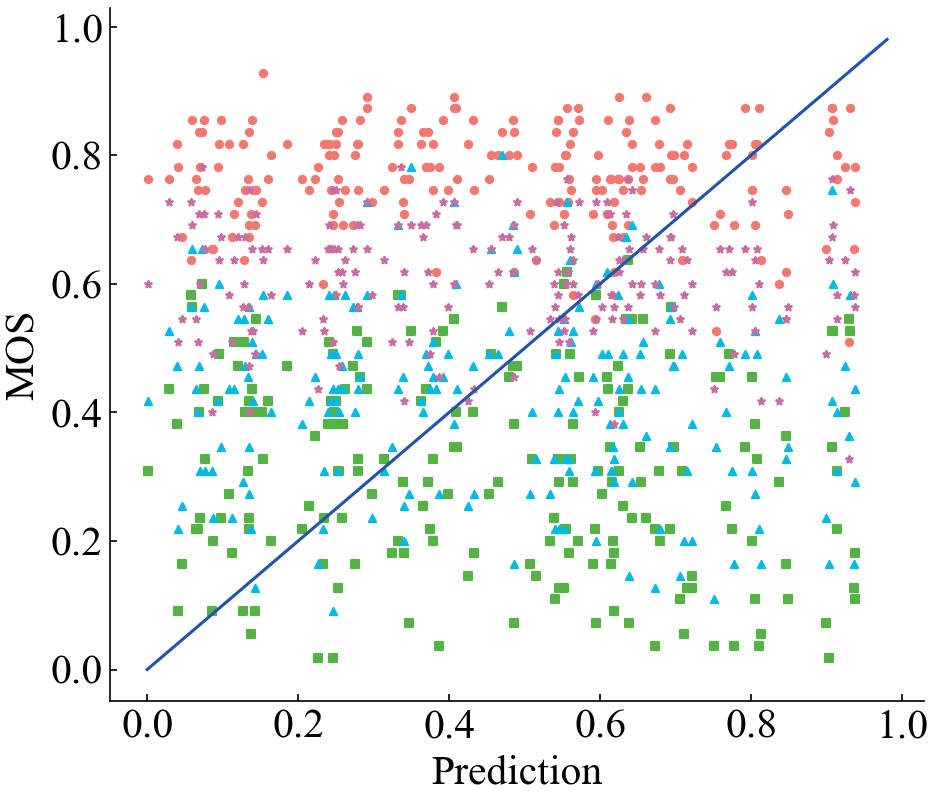}
    }
    \subfigure[CPP-PSNR]{
        \includegraphics[width=0.3\textwidth]{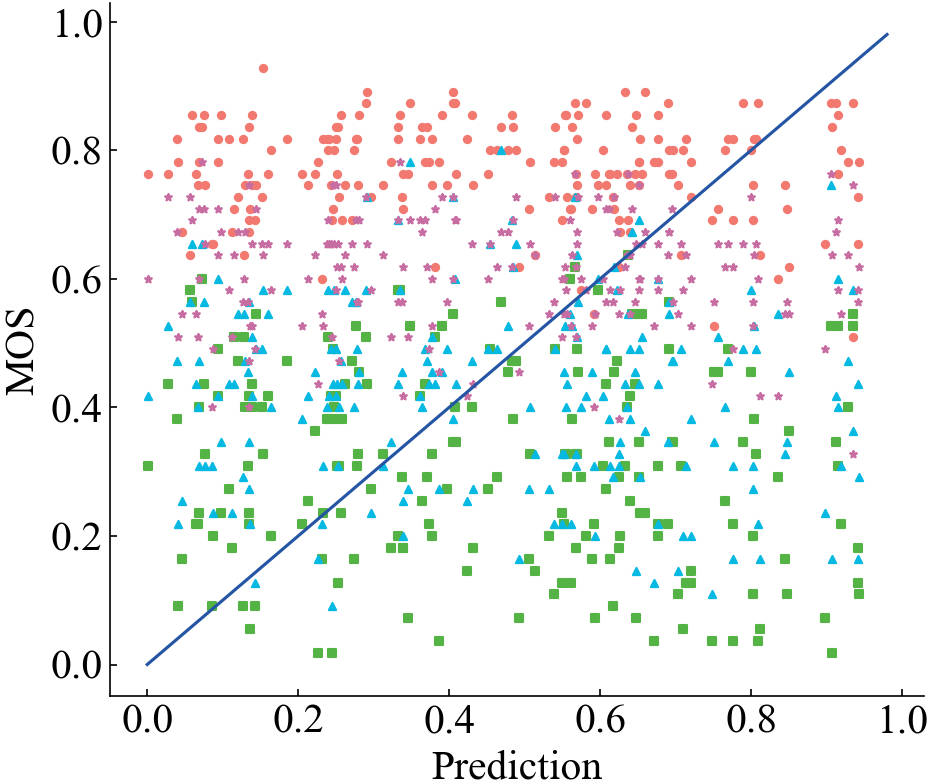}
    }
    
    \subfigure[WS-SSIM]{
        \includegraphics[width=0.3\textwidth]{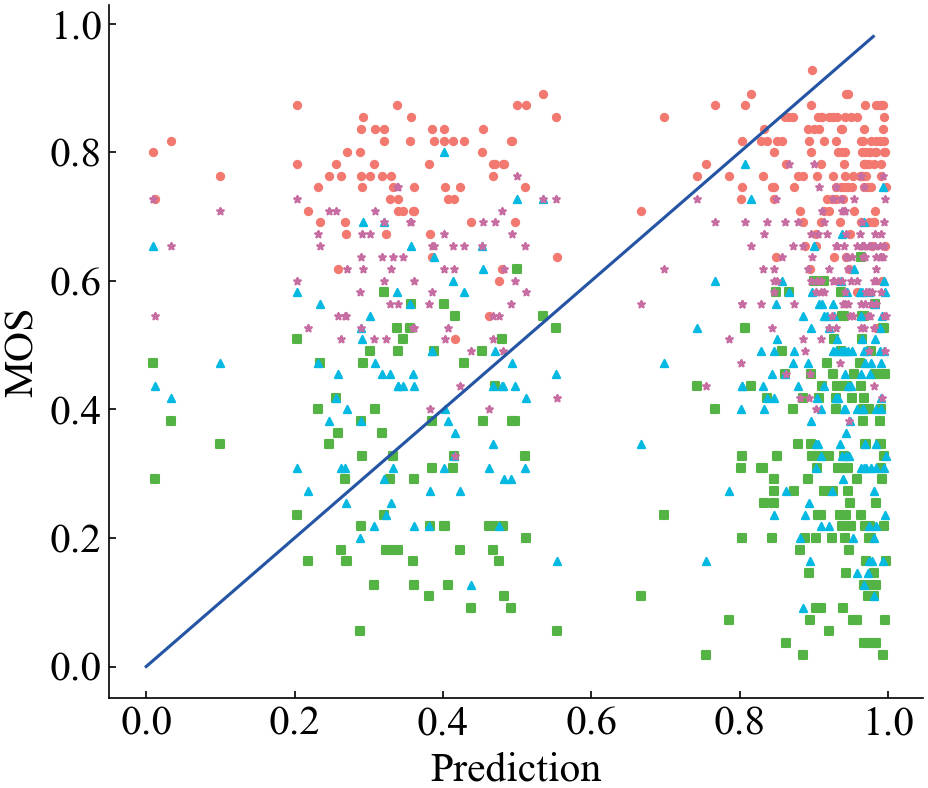}
    }
    \subfigure[MC360IQA]{
        \includegraphics[width=0.3\textwidth]{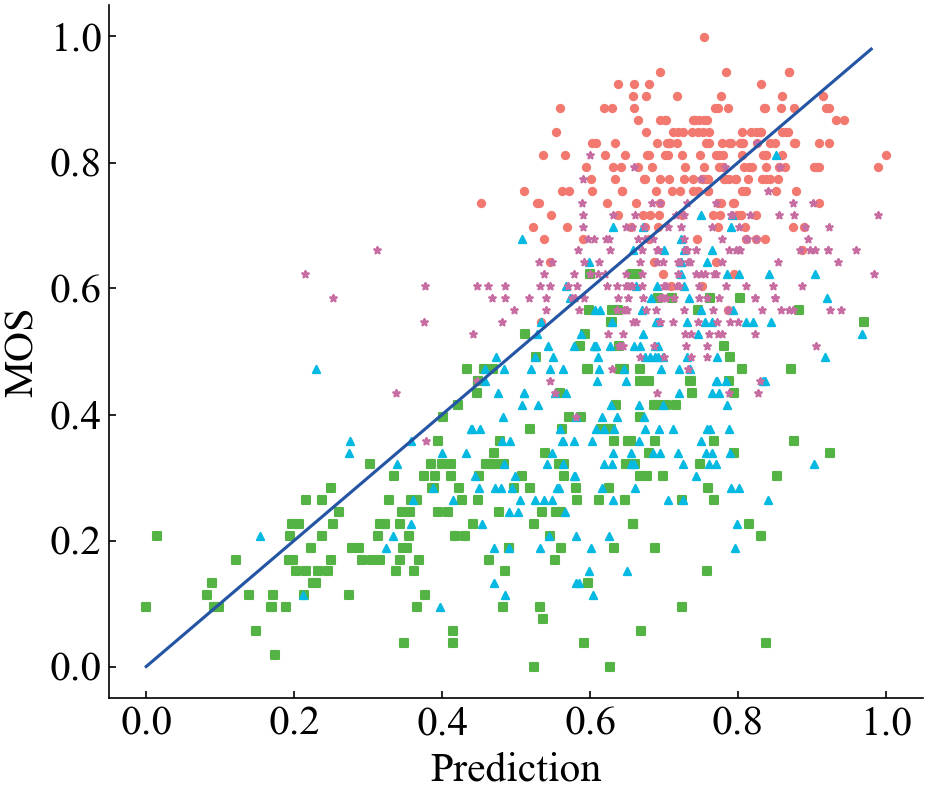}
    }
    \subfigure[VGCN]{
        \includegraphics[width=0.3\textwidth]{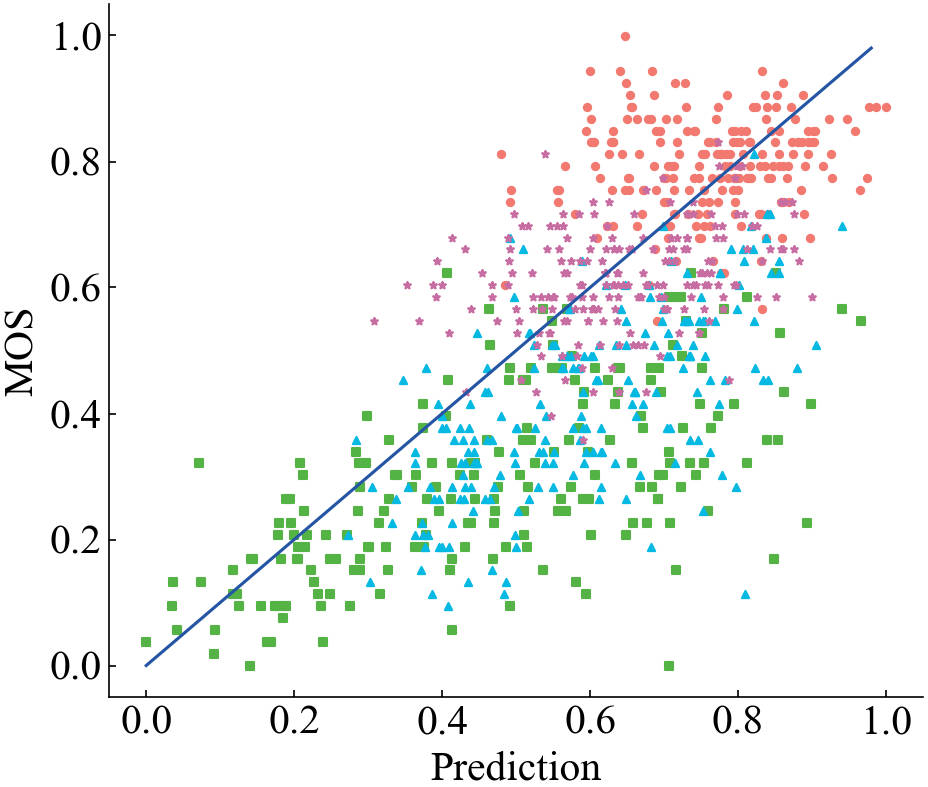}
    }
    
    \subfigure[Fang22]{
        \includegraphics[width=0.3\textwidth]{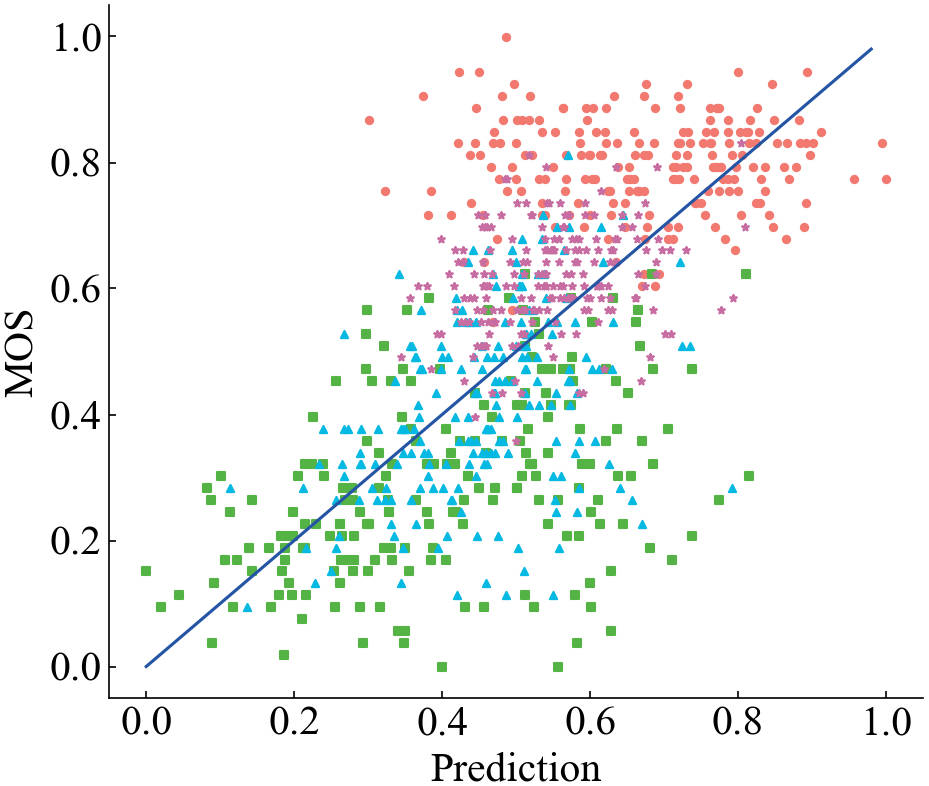}
    }
    \subfigure[Assessor360]{
        \includegraphics[width=0.3\textwidth]{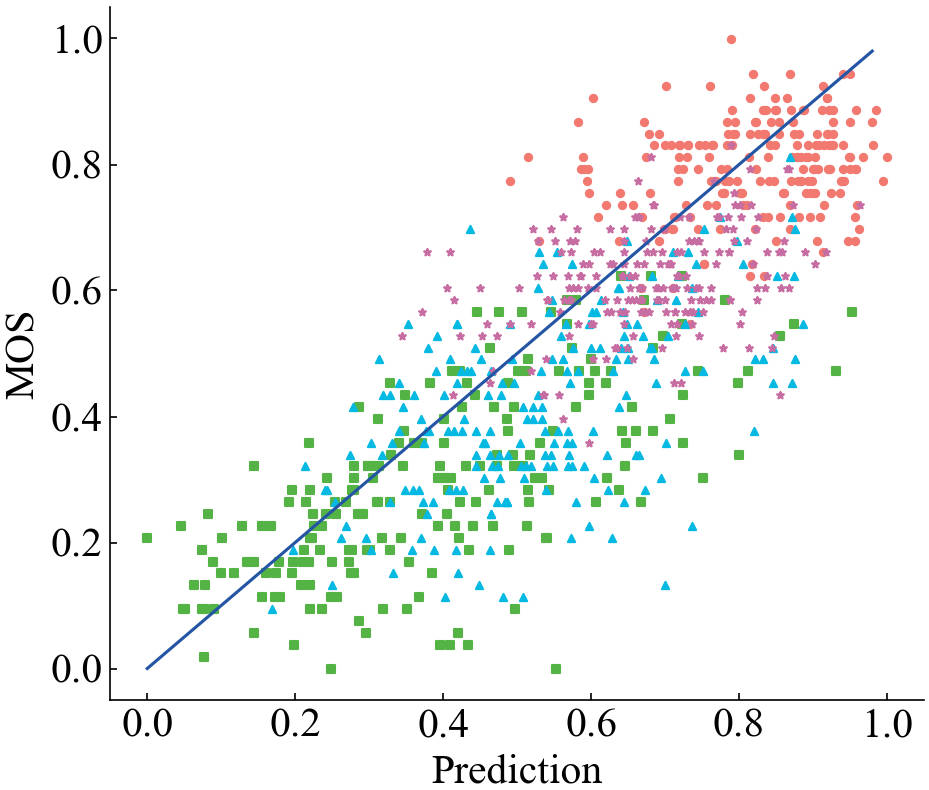}
    }
    \subfigure[Max360IQ]{
        \includegraphics[width=0.3\textwidth]{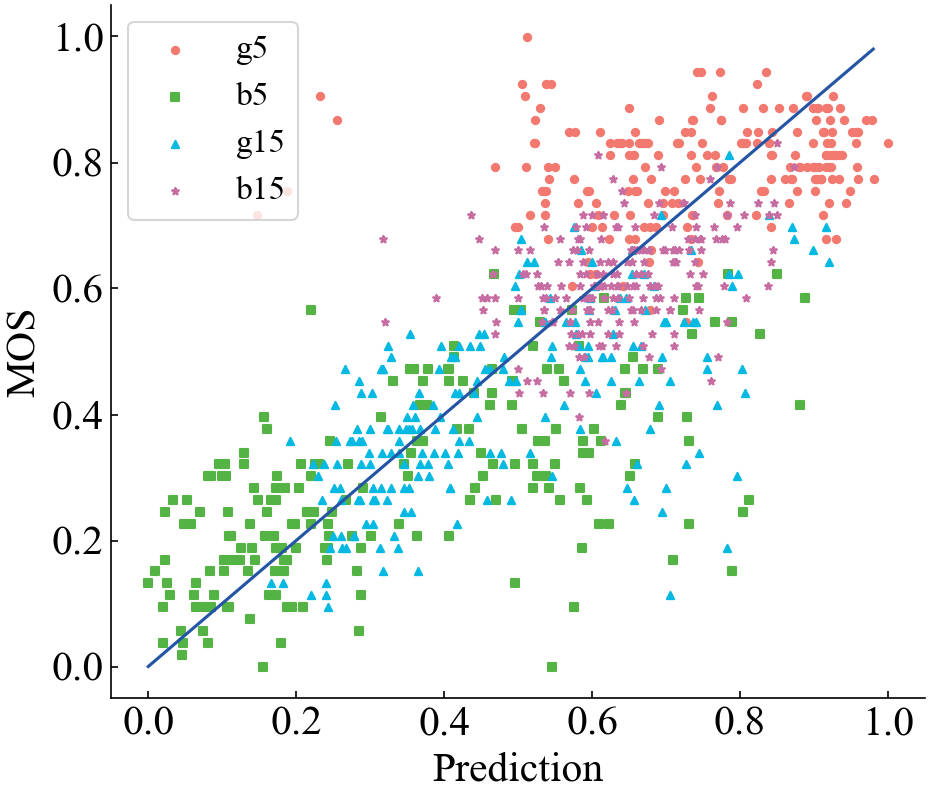}
    }
    
    \caption{The scatter plots of predictions by some methods against the subjective scores on the JUFE database. Note that the scatter plot for each color represents the predicted scores under different conditions.}
    \label{fig:models}
\end{figure}

We further study the performance of these test methods in different specific conditions on the JUFE database. The experimental results are shown in Tab.~\ref{tab:results2} and the scatter plots of subjective scores against predictions are illustrated in Fig.~\ref{fig:models}, where we have several interesting findings. First, it is noticeable that almost all methods obtain the best results in the Bad-5s condition but worse results in the Good-5s condition while the results of the Good-15s condition are better than Gad-15s condition. Second, for the condition with longer exploration time, Max360IQ and Assessor360 show better performance compared with other methods. We argue that the addition of GRUs effectively addresses the issue of the recency effect. Besides, the VGCN also shows impressive performance, which may be attributed to the utilization of GNN as the backbone, since the GNN is able to capture intricate spatial dependencies. However, all models still have a considerable gap from the optimal results, and there is much room for improvement in evaluating non-uniformly distorted OIs.

\begin{table*}[!htbp]
    \centering
    \caption{Performance comparison of state-of-the-art FR-OIQA and BOIQA methods on the JUFE database in detail. The best results are marked in \textbf{bold}.}
    \label{tab:results2}
    \begin{adjustbox}{max width=1\textwidth}
    \begin{tabular}{c|c|c|c|c|c|c|c|c|c|c|c|c|c|c|c|c}
        \toprule
        Type & Metrics & \multicolumn{3}{|c|}{Good-5s} & \multicolumn{3}{|c|}{Bad-5s} & \multicolumn{3}{|c|}{Good-15s} & \multicolumn{3}{|c|}{Bad-15s} & \multicolumn{3}{|c}{Average}\\
        \midrule
        &&PLCC&SRCC&RMSE&PLCC&SRCC&RMSE&PLCC&SRCC&RMSE&PLCC&SRCC&RMSE&PLCC&SRCC&RMSE\\
        \midrule
        \multirow{4}*{FR-OIQA} 
        & S-PSNR & 0.0532 & 0.0560 & 0.2830 & 0.1236 & 0.1108 & 0.5325 & 0.1057 & 0.0697 & 0.5124 & 0.0649 & 0.0720 & 0.3022 & 0.0869 & 0.0771 & 0.4075\\
        & WS-PSNR & 0.0554 & 0.0591 & 0.2830 & 0.1275 & 0.1127 & 0.5322 & 0.1101 & 0.1036 & 0.5122 & 0.0649 & 0.0739 & 0.3022 & 0.0895 & 0.0873 & 0.4074\\
        & CPP-PSNR & 0.0548 & 0.0549 & 0.2830 & 0.1258 & 0.1114 & 0.5324 & 0.1070 & 0.0695 & 0.5124 & 0.0665 & 0.0761 & 0.3021 & 0.0885 & 0.0780 & 0.4075\\
        & WS-SSIM & 0.0669 & 0.0365 & 0.2828 & 0.0952 & 0.0813 & 0.5342 & 0.0830 & 0.0848 & 0.5135 & 0.0575 & 0.0596 & 0.3023 & 0.0757 & 0.0656 & 0.4082 \\
        \midrule
        \multirow{4}*{BOIQA} 
        & MC360IQA & 0.2635 & 0.2186 & 0.2581 & 0.5772 & 0.5873 & 0.4197 & 0.4516 & 0.4334 & 0.4575 & 0.2681 & 0.2252 & 0.2757 & 0.3901 & 0.3661 & 0.3528\\
        & VGCN & 0.1680 & 0.1129 & 0.2637 & 0.6462 & 0.6429 & 0.3922 & 0.6371 & \textbf{0.6346} & 0.3952 & 0.3495 & 0.3154 & 0.2682 & 0.4502 & 0.4264 & 0.3298\\
        & Fang22 & 0.1384 & 0.1269 & 0.2650 & 0.4434 & 0.4376 & 0.4607 & 0.4291 & 0.3906 & 0.4631 & 0.2523 & 0.2423 & 0.2770 & 0.3158 & 0.2994 & 0.3665\\
        & ST360IQ & 0.2387 & 0.2436 & 0.2785 & 0.2303 & 0.2366 & \textbf{0.2729} & 0.2348 & 0.2281 & 0.4991 & 0.1664 & 0.1648 & 0.2932 & 0.2176 & 0.2183 & 0.3359\\
        & Assessor360 & 0.2698 & 0.2016 & 0.2576 & \textbf{0.6865} & \textbf{0.6982} & 0.3737 & 0.5470 & 0.5194 & 0.4292 & 0.3442 & 0.3147 & 0.2687 & 0.4619 & 0.4335 & 0.3323\\
        & Max360IQ & \textbf{0.3020} & \textbf{0.2692} & \textbf{0.2550} & 0.6477 & 0.6319 & 0.3916 & \textbf{0.6468} & 0.6249 & \textbf{0.3910} & \textbf{0.3904} & \textbf{0.3522} & \textbf{0.2635} & \textbf{0.4967} & \textbf{0.4696} & \textbf{0.3253}\\
        \bottomrule
    \end{tabular}
    \end{adjustbox}
\end{table*}

\subsection{Ablation studies}

We first prove the effectiveness of the designed MSFI module and DSG module, and the ablation experimental results are shown in Tab.~\ref{tab:ablation1}. We can find that the multi-scale quality-aware features hold greater significance when compared to the deep semantic features. This observation aligns with the human perception process~\cite{su2020blindly,wu2020end,yan2020blind}. Furthermore, by concatenating the two features, we can further improve the performance of our method. Besides, we explore the importance of GRUs when the viewport sequence has temporal features. The ablation experimental results are shown in Tab.~\ref{tab:ablation2}. From Tab.~\ref{tab:ablation2}, we can find that the GRUs are necessary for modeling the temporal dependency. We also discuss the influence of the backbone with different structures on the modeling of non-uniform and uniform distortion, three structures are taken into consideration, including CNN-based ResNet-50~\cite{he2016deep}, Transformer-based Swin Transformer~\cite{liu2021swin}, and hybrid-based MaxViT~\cite{tu2022maxvit}, the experimental results are shown in Tab.~\ref{tab:ablation_backbone}. The results prove that the hybrid-based MaxViT can effectively process both non-uniform and uniform distortion simultaneously with fewer parameters than the other two.

\begin{table}[!htbp]
    \centering
    \caption{Ablation study about each component in the Max360IQ on the JUFE database. The best results are marked in \textbf{bold}.}
    \label{tab:ablation1}
    \begin{adjustbox}{max width=0.4\textwidth}
    \begin{tabular}{c|c|c|c}
        \toprule
        Method & PLCC & SRCC & RMSE\\
        \midrule
        \multirow{3}*{} 
        DSG & 0.4141 & 0.4113 & 0.3424\\
        MSFI & 0.4817 & 0.4693 & 0.3285\\
        MSFI+DSG & \textbf{0.4967} & \textbf{0.4696} & \textbf{0.3253}\\
        \bottomrule
    \end{tabular}
    \end{adjustbox}
\end{table}

\begin{table}[!htbp]
    \centering
    \caption{Ablation study about GRUs on the JUFE database. The best results are marked in \textbf{bold}.}
    \label{tab:ablation2}
    \begin{adjustbox}{max width=0.5\textwidth}
    \begin{tabular}{c|c|c|c}
        \toprule
        Method & PLCC & SRCC & RMSE\\
        \midrule
        \multirow{2}*{} 
        backbone & 0.4321 & 0.4256 & 0.3440\\
        backbone+GRUs & \textbf{0.4967} & \textbf{0.4696} & \textbf{0.3253}\\
        \bottomrule
    \end{tabular}
    \end{adjustbox}
\end{table}

\begin{table}[!htbp]
    \centering
    \caption{Ablation study about various backbone. The best results are marked in \textbf{bold}.}
    \label{tab:ablation_backbone}
    \begin{adjustbox}{max width=1\textwidth}
    \begin{tabular}{c|c|c|c|c|c|c|c|c|c|c}
\toprule
Method & Params (M) &\multicolumn{3}{c|}{JUFE} & \multicolumn{3}{c|}{OIQA} & \multicolumn{3}{c}{CVIQ} \\
\midrule
  & wo/w GRUs & PLCC & SRCC & RMSE & PLCC & SRCC & RMSE & PLCC & SRCC & RMSE \\
\midrule
  ResNet-50 & 24.9 / 32.8 & 0.4720 & 0.4525 & 0.3344 & 0.9357 & 0.9389 & 0.7183 & 0.9832 & 0.9791 & 2.4340 \\
  Swin & 19.9 / 27.8 & 0.3108 & 0.2901 & 0.3696 & 0.9527 & 0.9519 & 0.6188 & 0.9797 & 0.9728 & 2.6752 \\
  MaxViT & 6.2 / 14.0 & \textbf{0.4967} & \textbf{0.4696} & \textbf{0.3253} & \textbf{0.9699} & \textbf{0.9704} & \textbf{0.4959} & \textbf{0.9844} & \textbf{0.9809} & \textbf{2.3485} \\
\bottomrule
    \end{tabular}
    \end{adjustbox}
\end{table}

Besides, we investigate the influence of the number of input viewports on the performance of the proposed Max360IQ. We find that the performance of the proposed Max360IQ on the JUFE database improves with the increase of viewports, and it begins to decline when the number of input viewports keeps increasing. This result implies that more viewports might introduce redundant information, and therefore potentially cause disturbance when the network is trained. By contrast, fewer viewports lead to better performance on the OIQA and CVIQ databases, due to the characteristic of uniform distortion that the quality of extracted viewports is close to the entire OI.

\begin{figure}[htbp]
    \centering

    \subfigure[JUFE]{
        \includegraphics[width=0.3\textwidth]{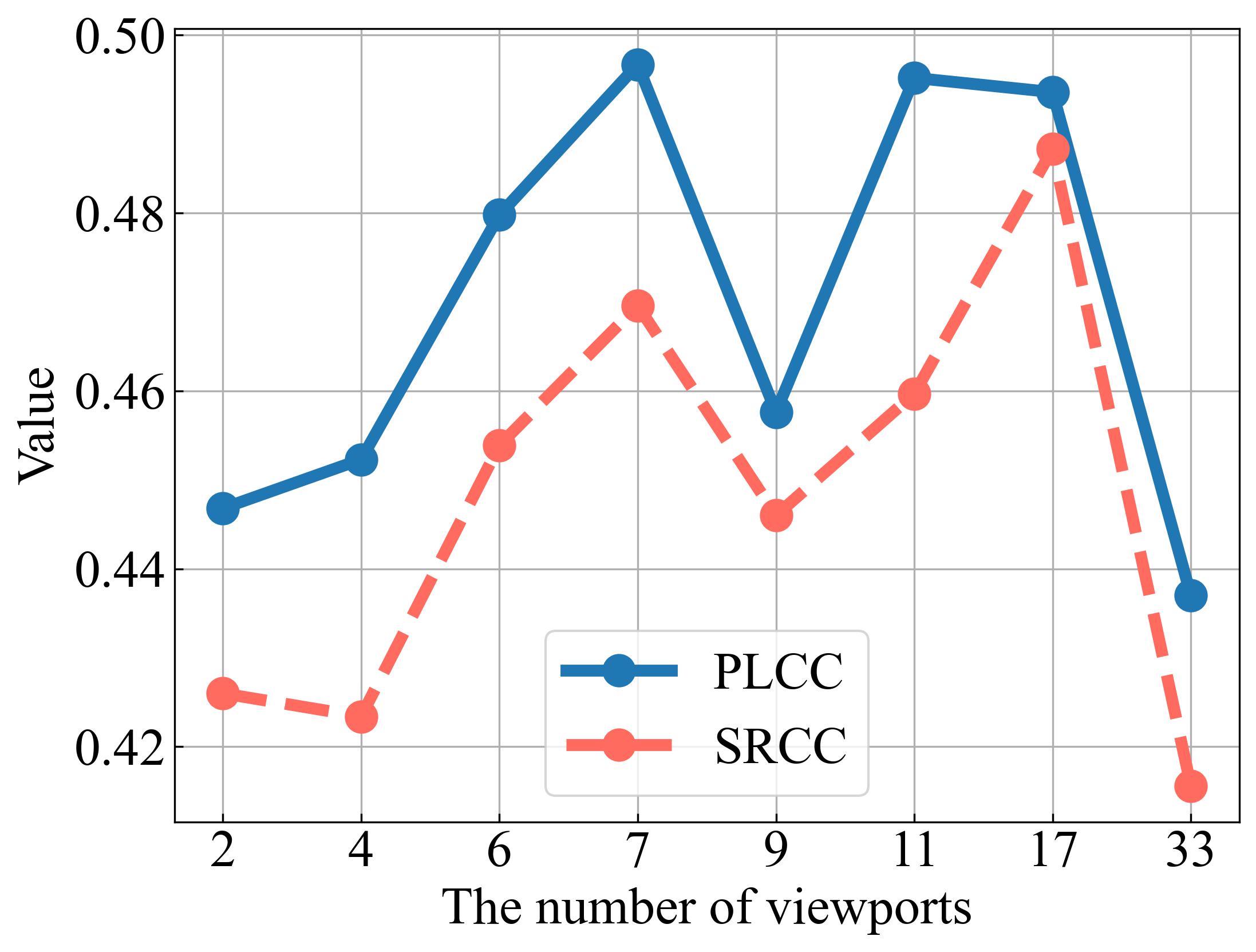}
    }
    \subfigure[OIQA]{
        \includegraphics[width=0.3\textwidth]{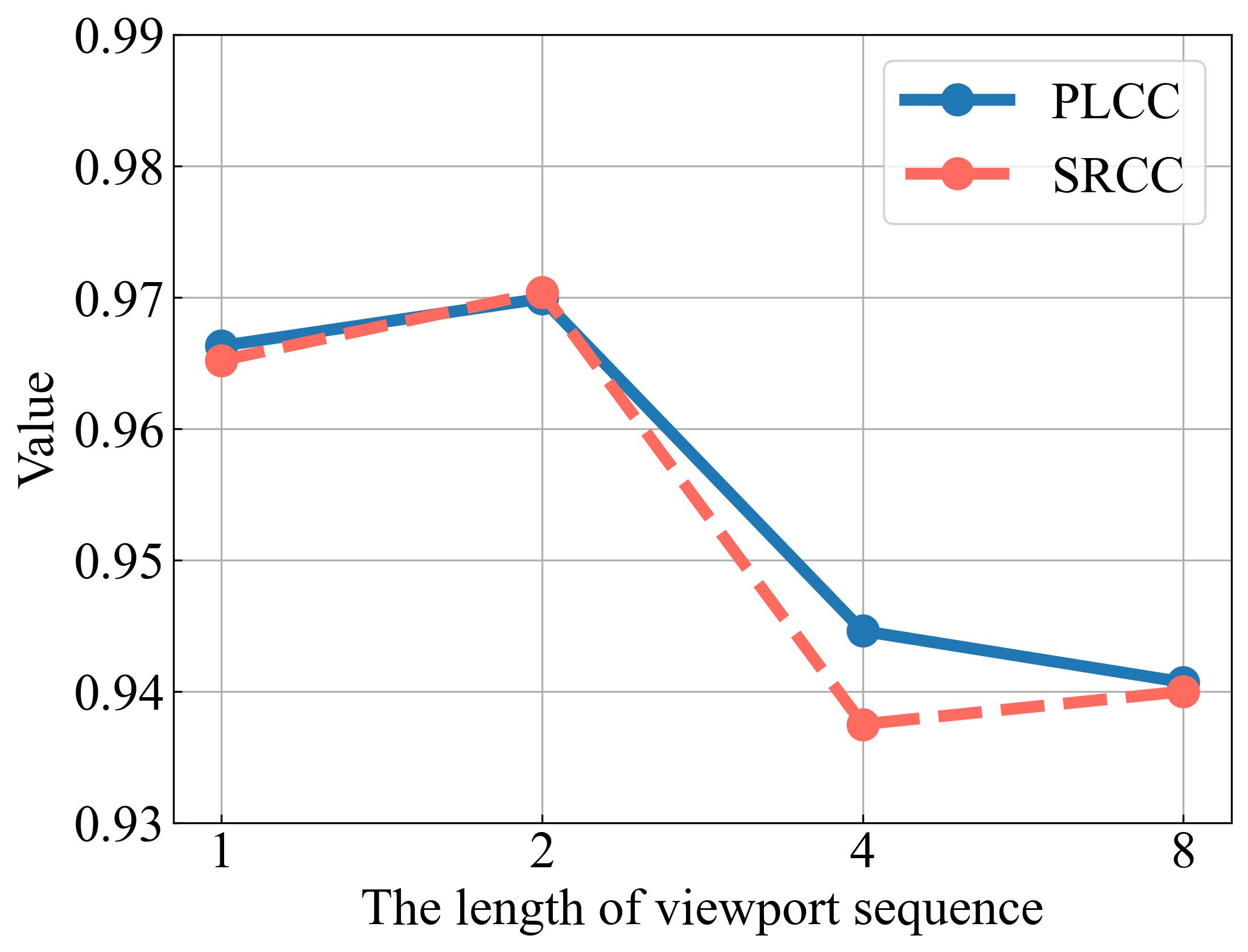}
    }
    \subfigure[CVIQ]{
        \includegraphics[width=0.3\textwidth]{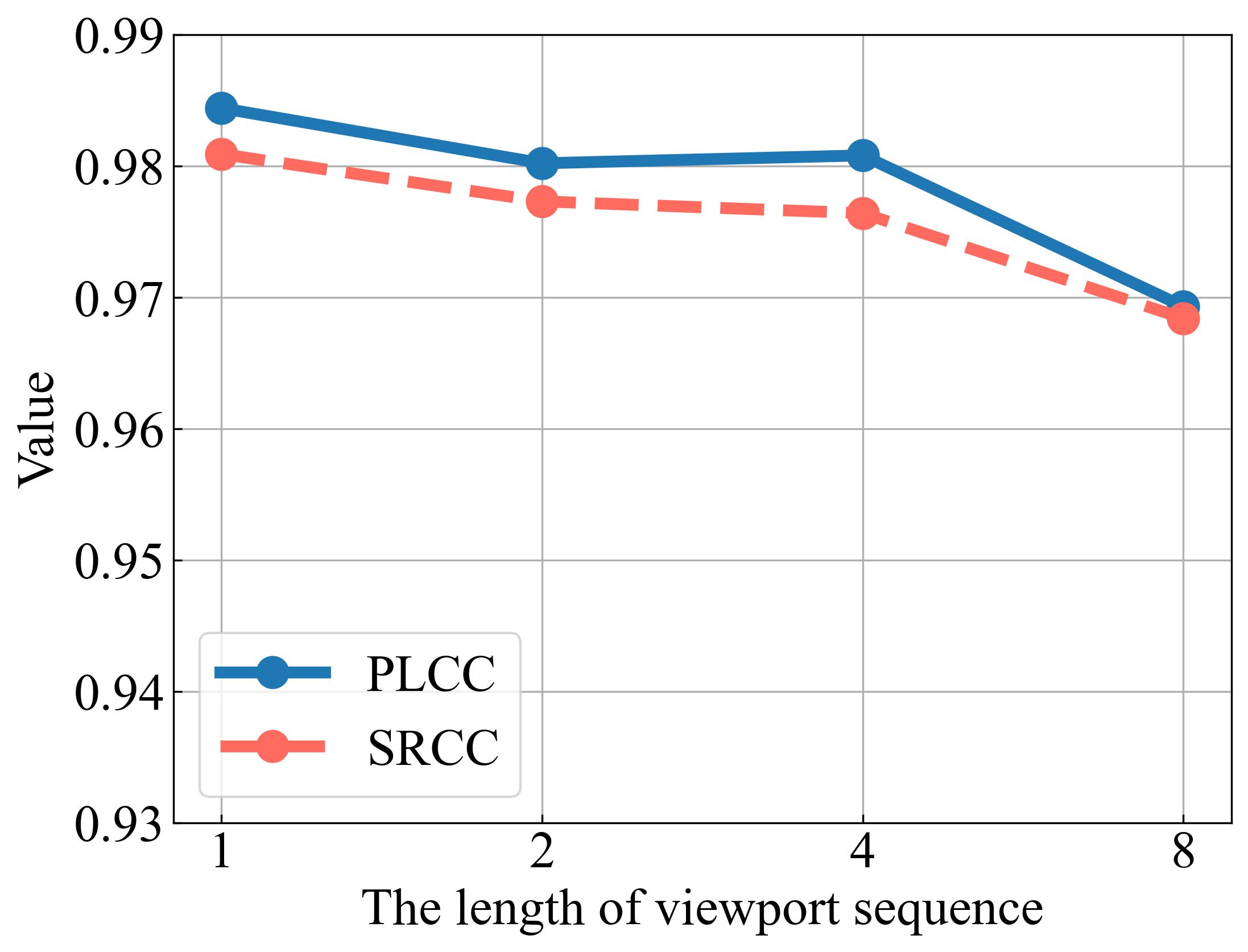}
    }
    \caption{The influence of the number of viewports on the performance of the proposed Max360IQ.}
    \label{tab:ablation3}
\end{figure}

\subsection{Visualization on Viewports}
\begin{figure}[htbp]
    \centering
    \subfigure[JUFE]{
        \includegraphics[width=0.3\textwidth]{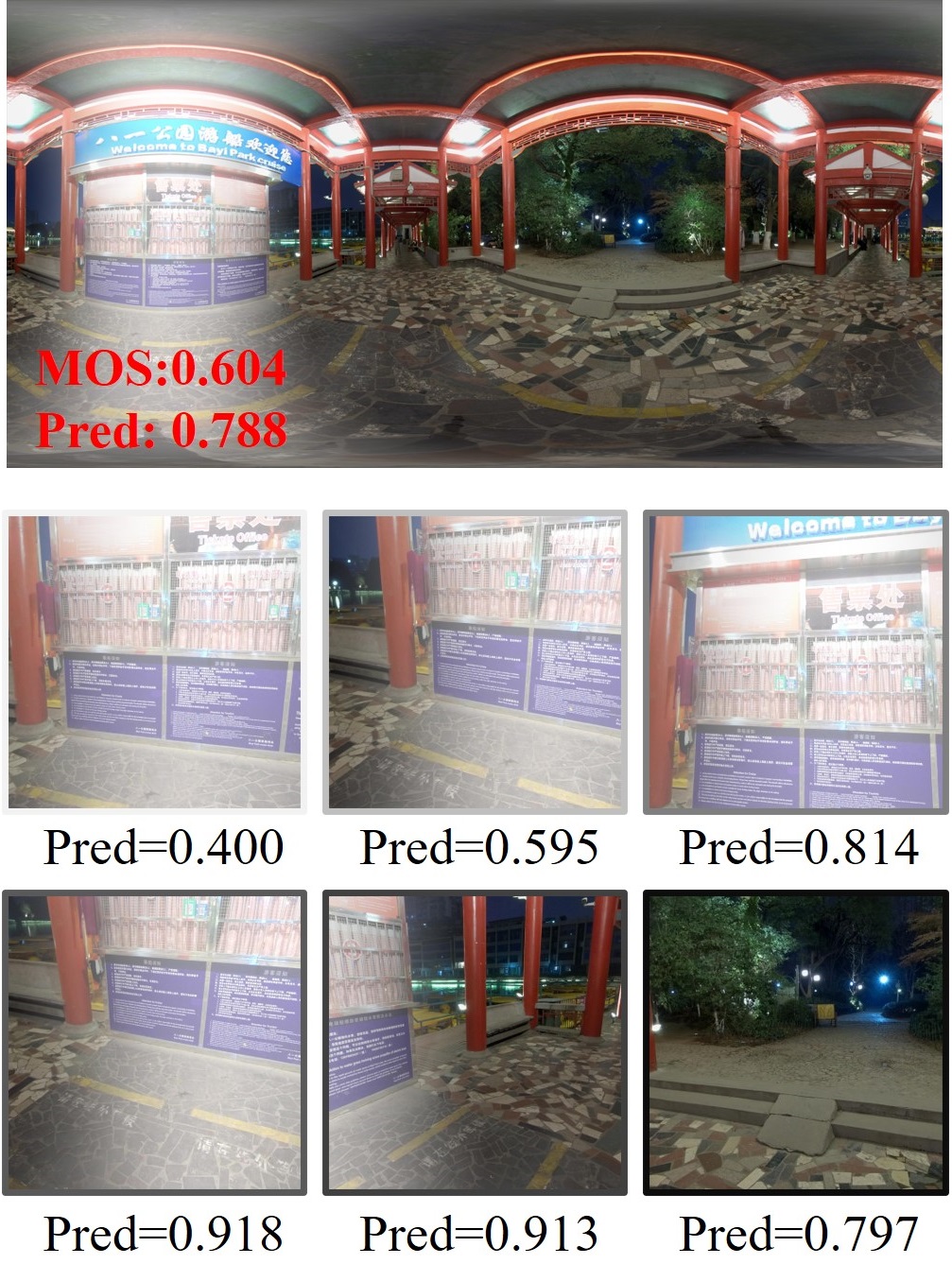}
    }
    \subfigure[OIQA]{
        \includegraphics[width=0.3215\textwidth]{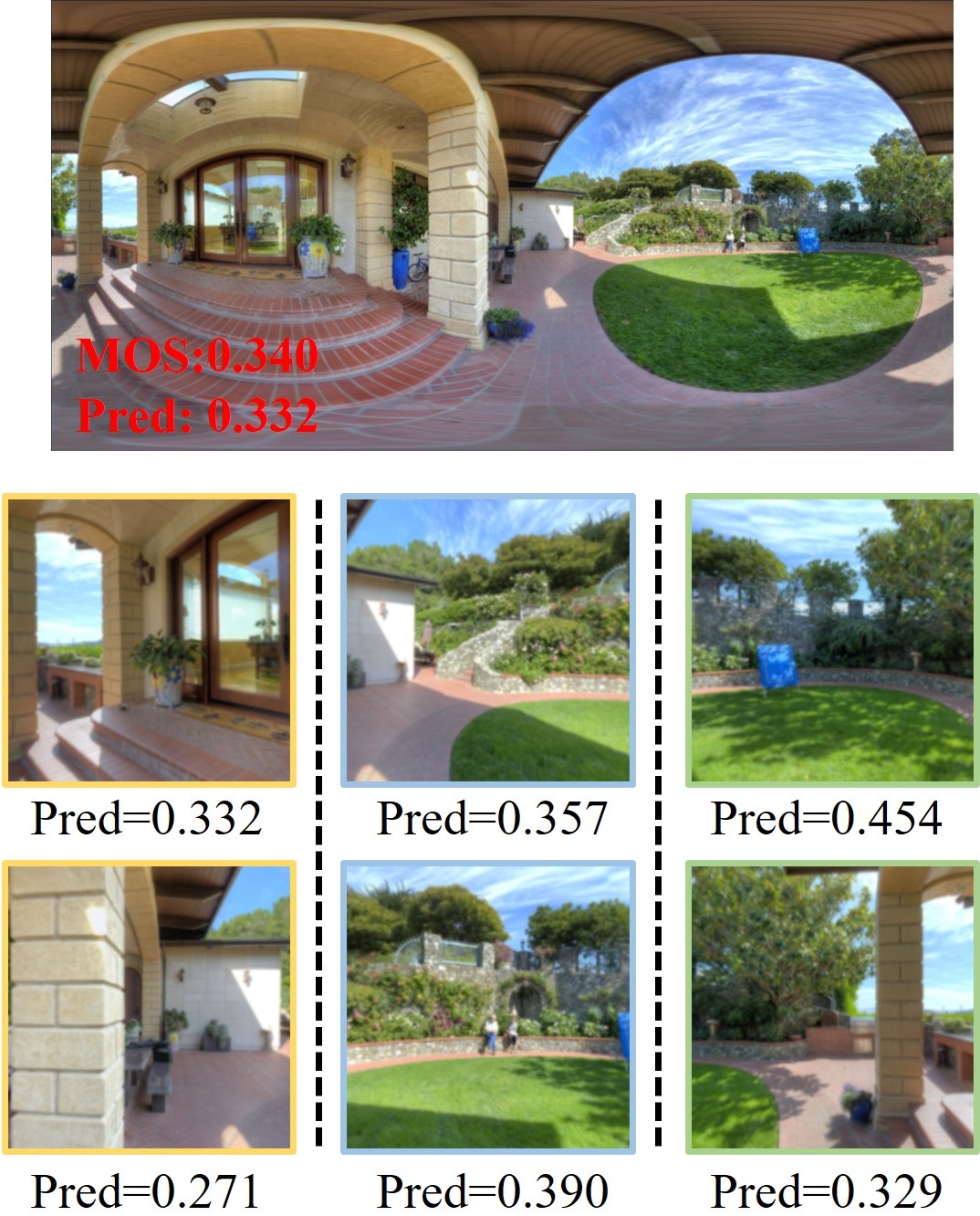}
    }
    \subfigure[CVIQ]{
        \includegraphics[width=0.3\textwidth]{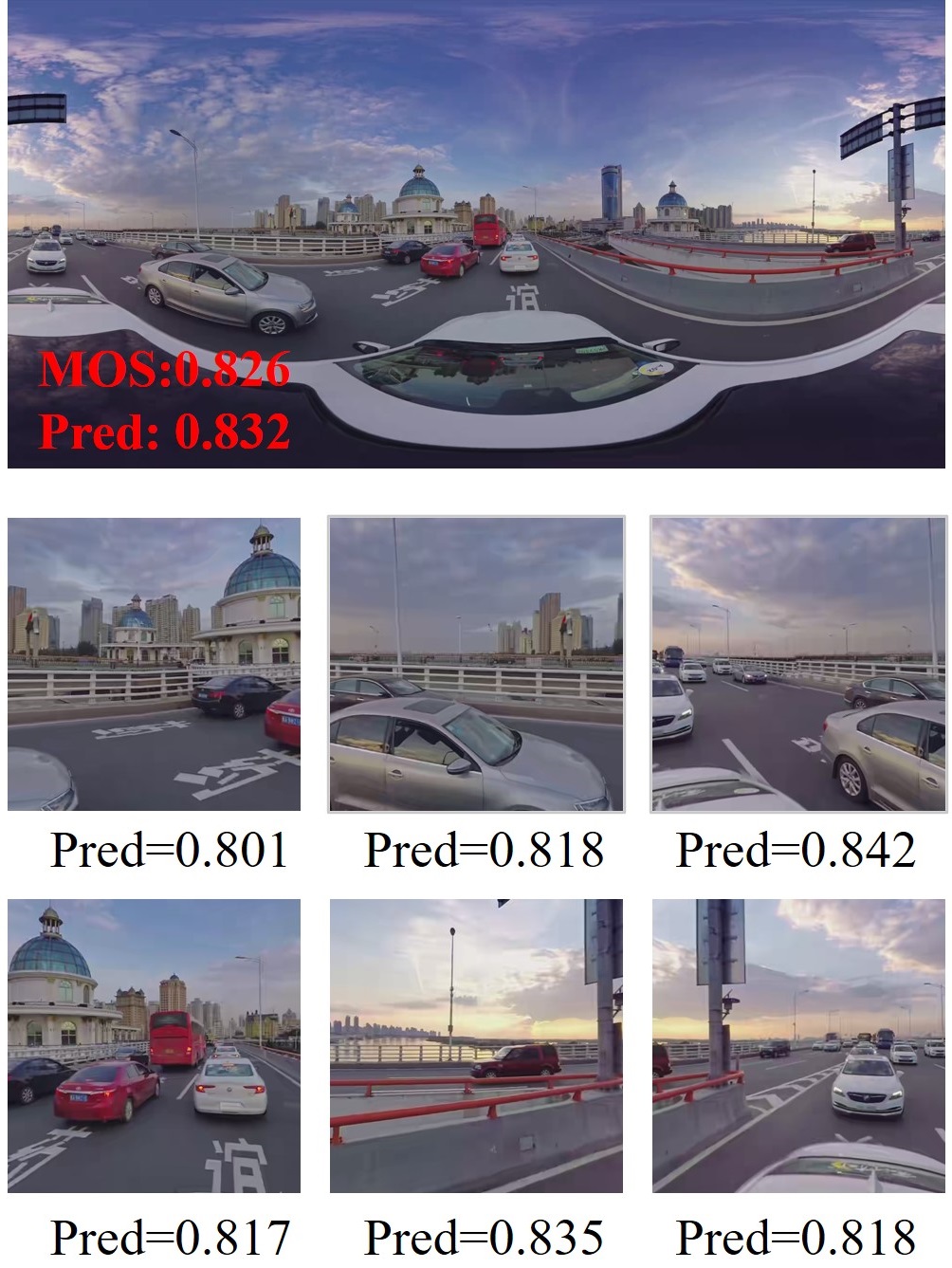}
    }
    \caption{The visual examples of the predicted results from the proposed model. Note that the viewport with a lighter contour is early viewed in (a), and the viewports with the same contour belong to one viewport sequence in (b).}
    \label{fig:visual}
\end{figure}

To objectively discuss the advantages and limitations of the proposed Max360IQ, Fig.~\ref{fig:visual} visualizes the predicted results of several examples and their viewports. For the non-uniformly distorted OIs in the JUFE database, the proposed model successfully tackles the recency effect in user viewing behavior. However, the prediction of the proposed model is not sufficiently precise, because there still remains a challenge that how to effectively aggregate the varied local scores into an overall score rather than a simple average operation. For the uniformly distorted OIs in the OIQA and CVIQ databases, we observe that the predicted quality scores for individual viewports closely align with those of the entire OIs, meaning the high failure tolerance in the prediction through the average operation. Besides, this consistency also facilitates data augmentation during the training phase by solely utilizing one or two viewports as input, which improves the training process efficiently.

%-------------------------------------------------------------------------
\section{Conclusions}
In this paper, we propose a BOIQA model named Max360IQ based on multi-axis attention. First, we extract the viewport sequences of each OI using suitable extraction approach for IOs with non-uniform and uniform distortion. Then, we utilize stacked multi-axis attention modules as the backbone to capture global and local spatial interactions of viewports. To align with the HVS, we design a MSFI module to further fuse the interactions of viewports at different scales and devise a deep semantic guided quality regression module for mapping the extracted features to quality score. In the experiments, we find that the multi-scale quality perception is of great significance in BOIQA and the GRUs component is necessary to process the viewport sequence with temporal dependency, and the proposed Max360IQ outperforms the state-of-the-art Assessor360 by 3.5\% in terms of PLCC and 3.6\% in terms of SRCC on the JUFE database with non-uniform distortion,
and has improved on the OIQA and CVIQ databases with uniform distortion according. Our proposed model can be applied in various scenarios, such as automatic high-quality omnidirectional image selection, aiding omnidirectional photography and video production, and testing the performance of omnidirectional cameras and equipment. However, it is significant to note that there is no OIQA model that can remarkably well assess the quality of non-uniformly distorted OIs. Relevant research in this area remains to be explored and developed.

\bibliography{elsarticle}





% \end{thebibliography}
\end{document}